\documentclass[a4paper]{article}
\usepackage{cite}
\usepackage{wrapfig}
\usepackage{graphicx}
\usepackage{amssymb}
\usepackage{amsfonts}
\usepackage{amsmath}
\usepackage{longtable}
\usepackage{rotating}
\usepackage{lscape}
\usepackage{epsfig}
\usepackage{multirow}

\begin{document}
\begin{center}
{\large\bf{Influence of triaxiality on the description of low-energy excitation spectrum of $^{96}$Zr}}\\[3mm]
%\maketitle

{\bf{E.\,V.\,Mardyban$^{a,b,}$\footnote{E-mail: mardyban@theor.jinr.ru},
T.\,M.\,Shneidman$^{a}$, E.\,A.\,Kolganova$^{a,b}$, R.\,V\,Jolos$^{a,b}$}}\\[3mm]

\noindent
{$^{a}$\,Joint Institute for Nuclear Research, 141980 Dubna, Moscow region, Russia}\\
{$^{b}$\,Dubna State University, 141982 Dubna, Moscow Region, Russia}\\
%\from{$^{c}$\,Kazan Federal University, Kazan 420008, Russia}
\end{center}
\begin{abstract}
The observed properties of the low-lying collective states of $^{96}$Zr  are investigated
within the geometrical collective model. The quadrupole-collective Bohr Hamiltonian with the potential having spherical  and axially-symmetric deformed minima is applied. The role of triaxiality is investigated by rotating the potential  in $\gamma$-degree of freedom so that the deformed minimum occurs at various axially  asymmetric shapes.  The change of excitation energies and  reduced matrix elements of quadrupole transitions with increase of triaxiliaty is analyzed.
\end{abstract}
\vspace*{6pt}


 \section{Introduction}
  The  gradual  transition from spherical to deformed shapes is known to occur in isotopic chains of many nuclei. 
 The notable feature of nuclear structure in A $\approx$ 100 nuclei, and in Zr isotopes in particular,  is an abrupt change of shape  \cite{1,3}. 
  In the previous work \cite{Mardyban, Sazonov} it was shown that the low energy structure of $^{96}$Zr can be satisfactorily described within the framework of the geometric collective model using the Bohr Hamiltonian. The dependence of the potential energy on the parameters of the quadrupole deformation $ \beta $ and $\gamma$ was fixed for the best description of the experimental data on the excitation energies and reduced transition probabilities $B(E2;  2^+_2\rightarrow0^+_2)$, $B(E2; 2^+_1\rightarrow0^+_1)$ and $B(E2; 2^+_2\rightarrow0 ^ +_1) $ The resulting potential has  two minima - spherical and deformed, separated by a barrier, which explains the coexistence of forms found in this nucleus \cite{Heyde1, Garcia}. In \cite{Mardyban,Sazonov}, the deformed minimum was assumed to have an  axial symmetry.  It is interesting, however, to investigate  to which extend the assumption of axial symmetry is important to reproduce the experimental data on $^{96}$Zr.

\section{Model}
Following the considerations outlined in \cite{Mardyban}, the model Hamiltonian is:
\begin{eqnarray}
\label{Hamiltonian_simplified_2}
H&=& -\frac{\hbar^2}{2 B_0} \left (\frac{1}{\beta^4} \frac{\partial}{\partial \beta}\beta^4 \frac{\partial}{\partial \beta}
+\frac{1}{\beta^2 \sin{3\gamma}} \frac{\partial}{\partial \gamma}\sin{3\gamma}\frac{\partial}{\partial \gamma}\right. \nonumber \\
&+& \left. \sum_{k=1}^3 \frac{\hat J^2_k}{4 b_{rot} \beta^2 \sin{\left (\gamma - \frac{2 \pi k}{3}\right )}}\right )+V(\beta, \gamma),
\end{eqnarray}
where
\begin{eqnarray}
\label{brot}
b_{rot}=
        \begin{cases}
            1 & \text{if $\beta \le \beta_m$, } \\
            b_{def}<1 & \text{if $\beta > \beta_m$.}
        \end{cases}
\end{eqnarray}
The value of $b_{rot}$ is obtained by fitting the energy of the lowest $2^{+}$ state localized in deformed minimum, that is $2^+_2$ state. The transition from the spherical to the deformed value of $b_{rot}$ occurs at $\beta = \beta_m$, where $\beta_m$ is selected in the vicinity of the maximum of the barrier separating  spherical and deformed potential wells. Our calculations show that the exact value of $\beta_m$ does not affect the calculation results.

The eigenfunctions of the Hamiltonian (\ref{Hamiltonian_simplified_2}) are obtained as a series expansion
\begin{eqnarray}
\Psi_{nIM}=\sum_i c^n_i R_i(\beta)\Upsilon^{I M}_i(\gamma, \Omega),
\label{basis}
\end{eqnarray}
where  $\Upsilon^{IM}_{n_\gamma}$ is the   SO(5)$\supset$SO(3) spherical harmonics and $R_{n_\beta}$ is  the eigenfunctions of the Hamiltonian of the five-dimensional harmonic oscillator.
The  construction of $\Upsilon^{I M}_i(\gamma, \Omega)$  is described in \cite{Bes1959, Rowe2004}. 
In Eq.~\eqref{basis},  the angular momentum and its projection on the laboratory axis are denoted as $I$  and $M$, respectively.

Using the obtained wave functions, the reduced matrix elements of the electric quadrupole transition are calculated. The collective quadrupole operator is taken in the standard form:
\begin{eqnarray}
\label{Q2operator}
Q^{coll}_{2 \mu}
=\frac {3Ze}{4\pi}R_0^{2} \left (\beta \cos{\gamma}D^{2}_{\mu 0}(\Omega) + \frac{1} {\sqrt{2}} \beta \sin{\gamma}\left ( D^{2}_{\mu 2}(\Omega)\right.\right.\nonumber\\
\left.\left. + D^{2}_{\mu -2}(\Omega)\right )\right ),
\end{eqnarray}
where $R_0$ is the radius of the equivalent spherical nucleus, and $Z$ is the charge number of the nucleus.

\section{Potential enegry}

In Ref.~\cite{Mardyban}, the potential energy $V(\beta, \gamma)$  was chosen in the form:
\begin{eqnarray}
V(\beta,\gamma)=U(\beta) +C_\gamma \beta^3(1-\cos{3\gamma}),
\label{potential energy}
\end{eqnarray}
to ensure that the deformed minimum  occurs at $\gamma = 0$. This form of potential energy  provides a weak $\gamma$-dependence at small $\beta$ due to the factor $\beta ^ 3$. The value $U(\beta)$ of the potential energy at $\gamma = 0$ and the parameter $C_\gamma $, which determines the rigidity of the potential relative to $\gamma$ in the deformed minimum, are chosen so as to reproduce the experimental data. Ultimately, $C_\gamma$ was fixed at 50 MeV in order to reproduce a reasonable value of the $\gamma$ vibration frequency, close to 1.5 MeV. The deformation at the second minimum was taken equal to $\beta=0.24$ in accordance with the experimental value of $B(E2; 2^+_2\rightarrow0^+_2)$.

\begin{figure*}[t]
%\begin{center}
\centering
\includegraphics[width=0.32\textwidth]{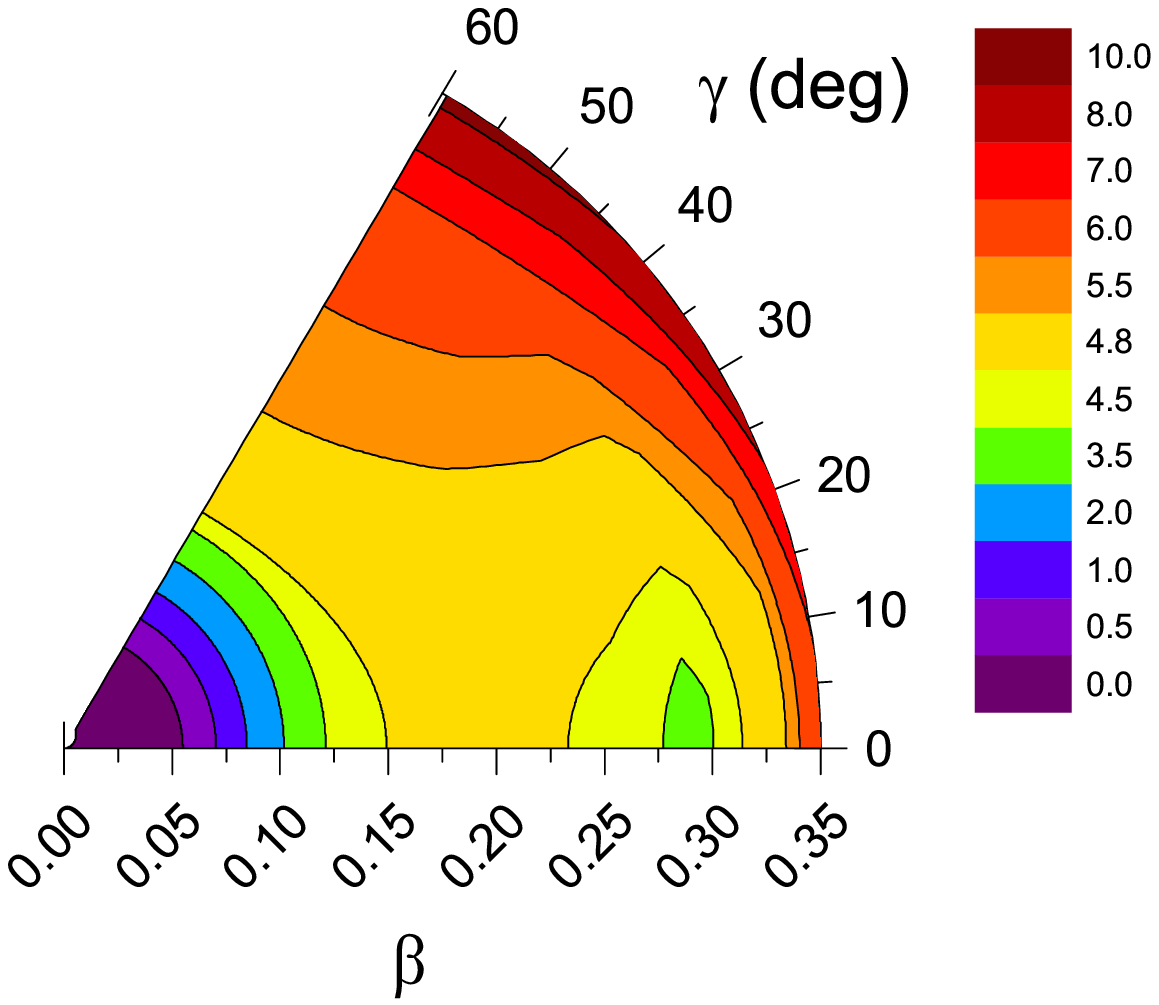}
\includegraphics[width=0.32\textwidth]{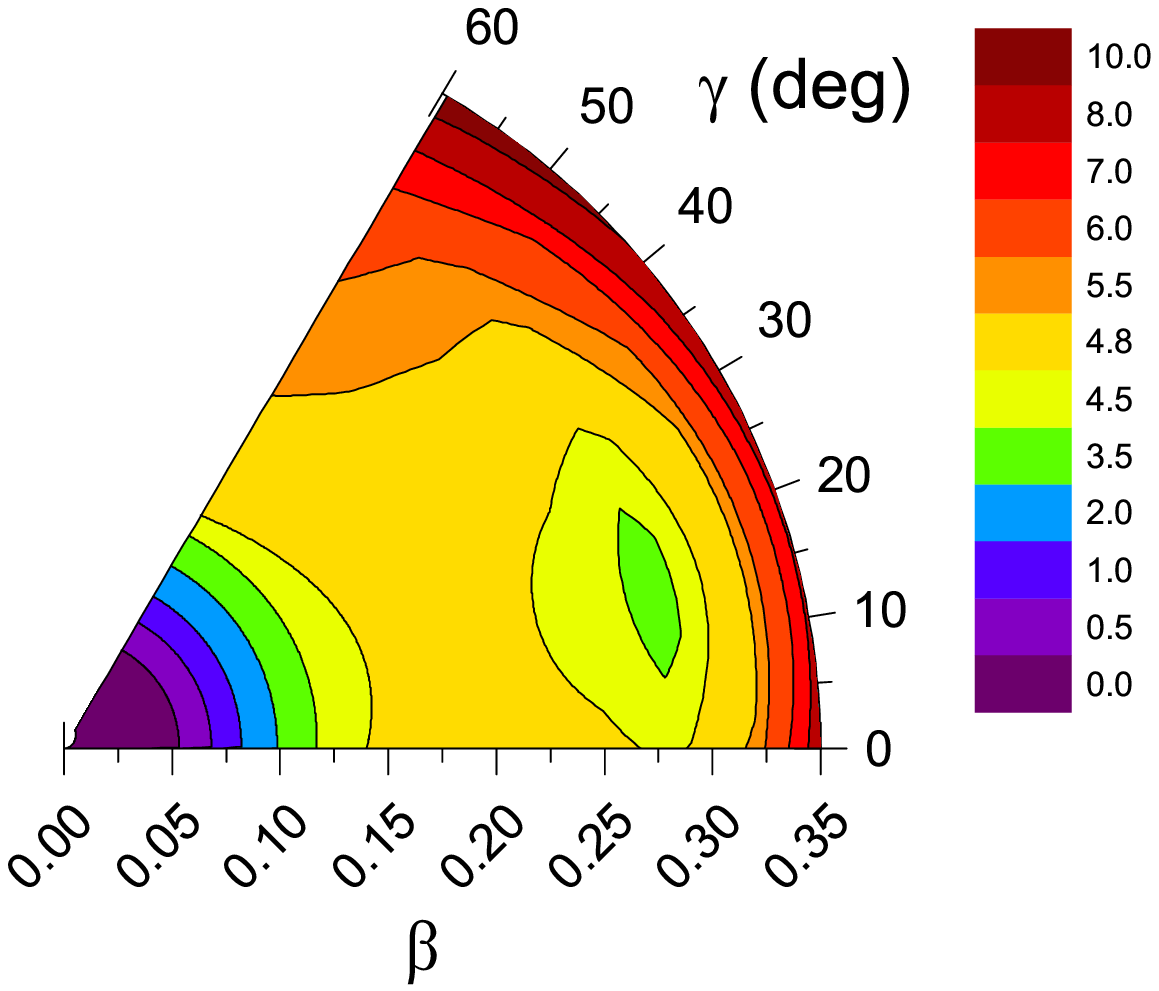}
\includegraphics[width=0.32\textwidth]{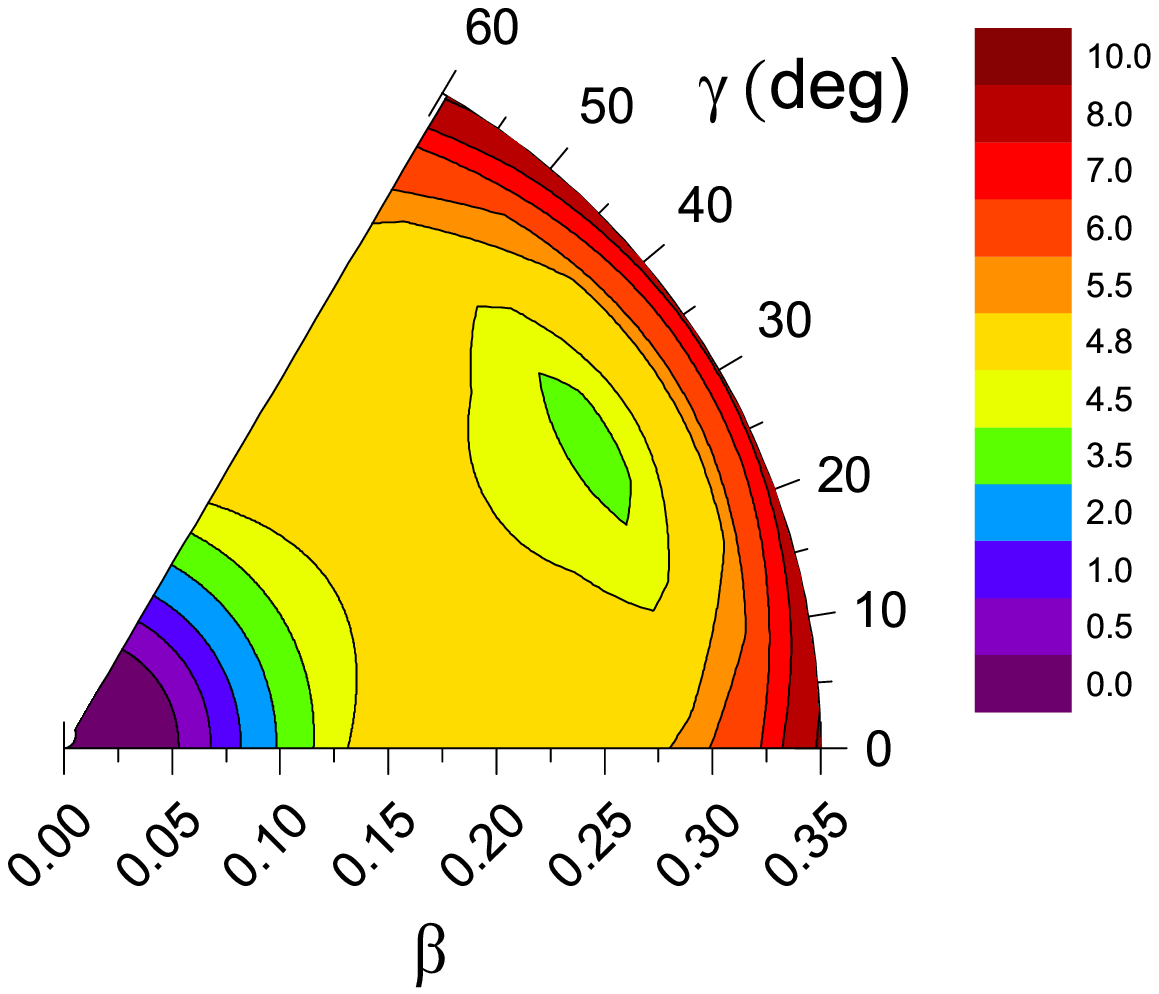}
%\end{center}
\caption{\label{f1} Potential energy $V (\beta, \gamma) $ with deformed minimum at $ \gamma = 0, 15$ and 30$^\circ$ respectively.}
\end{figure*}

In order to investigate how the description of the experimental data changes when the deformed minimum is shifted to the region of non-axial deformation, we calculate spectrum and transition probabilities with the potential 
\begin{eqnarray}
\tilde V(\beta,\gamma)=U(\beta) +C_\gamma \beta^3(1-\cos{\left (3[\gamma-\gamma_0]\right)}),
\label{potential energy2}
\end{eqnarray} 
obtained by the rotation  of the  potential  $V(\beta,\gamma)$ through the angle $\gamma_0$. 
Since  moments of inertia are assumed to be independent of deformation, the Hamiltonian (\ref{Hamiltonian_simplified_2}) is symmetric with respect to the change of  $\gamma_0$ to $\pi/3- \gamma_0$, where  $ 0\leqslant \gamma_0\leqslant \pi/6$. In particular, cases of prolate ($\gamma_0 = 0^\circ$) and oblate  ($\gamma_0 = 60^\circ$)  deformations are equivalent.  Thus, it is sufficient to diagonalize Hamiltonian (\ref{Hamiltonian_simplified_2}) with potential energy  $\tilde V(\beta,\gamma)$ only  for  $ 0\leqslant \gamma_0\leqslant \pi/6$.
As an example, the potential energy surfaces with a deformed minimum at $\gamma = 0^\circ, 15^{\circ}$ and $30^\circ$ are shown In Fig.~1.

\section{Results}
%\begin{center}

In Fig. 2 the dependences of the reduced probabilities of some electromagnetic transitions on the rotation angle $\gamma_0$ are shown. We see that with few exceptions, $B(E2)$'s weakly depend on the triaxiality of potential. For the transitions between states localized in spherical well this behaviour can be explained as due to fact that the region of the potential close to $\beta = 0 $ practically does not change upon rotation. Indeed, in Fig. 3 the weights of various $K$ components of wave functions of excited states are shown. We see that structure of $2^+_1$ and $4^+_1$ states almost independent on $\gamma_0$. Similarly the energies of these levels change smoothly (see Table~1).

\begin{figure}[t]
%\begin{center}
\centering
\includegraphics[width=0.32\textwidth]{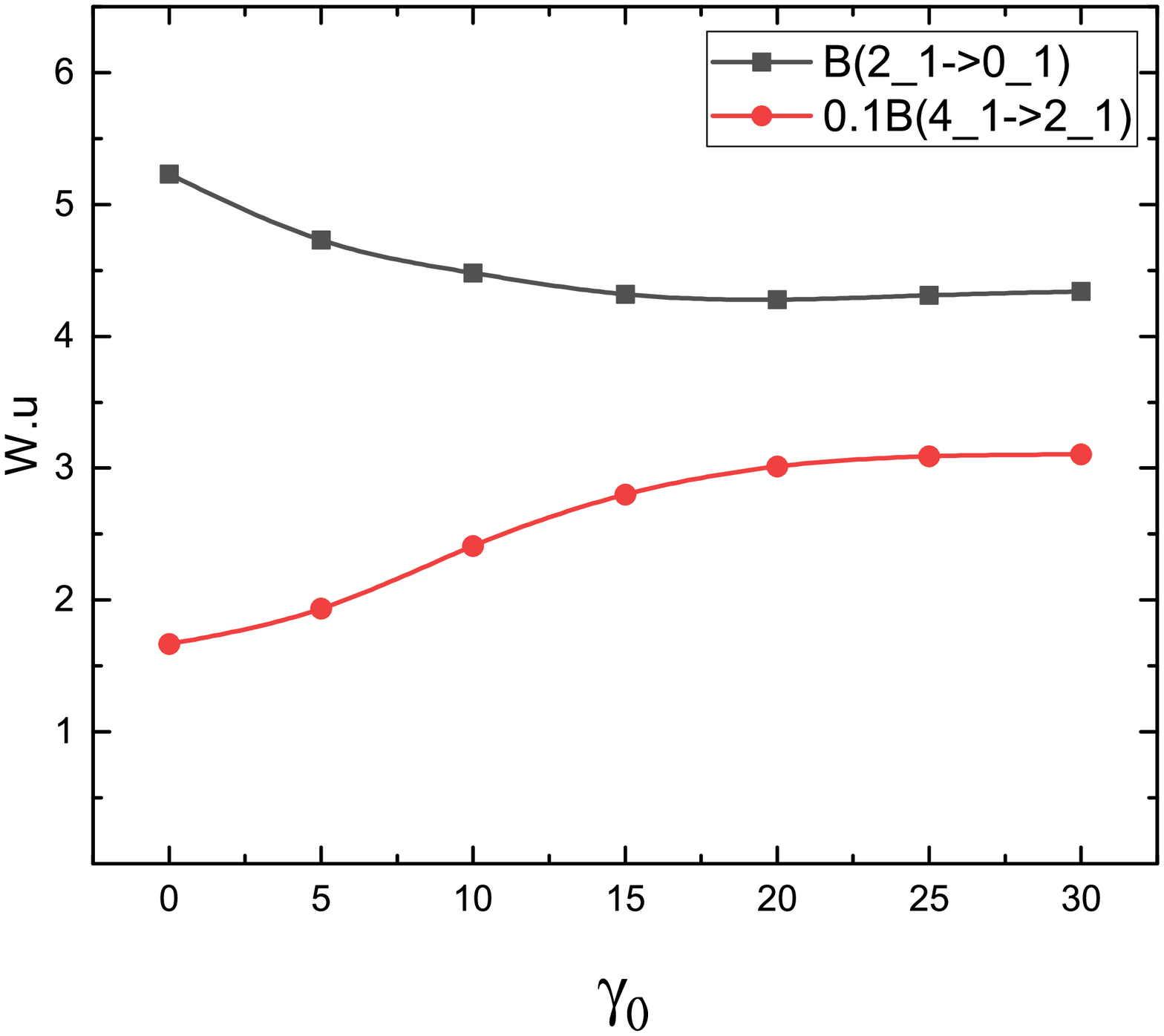}
\includegraphics[width=0.32\textwidth]{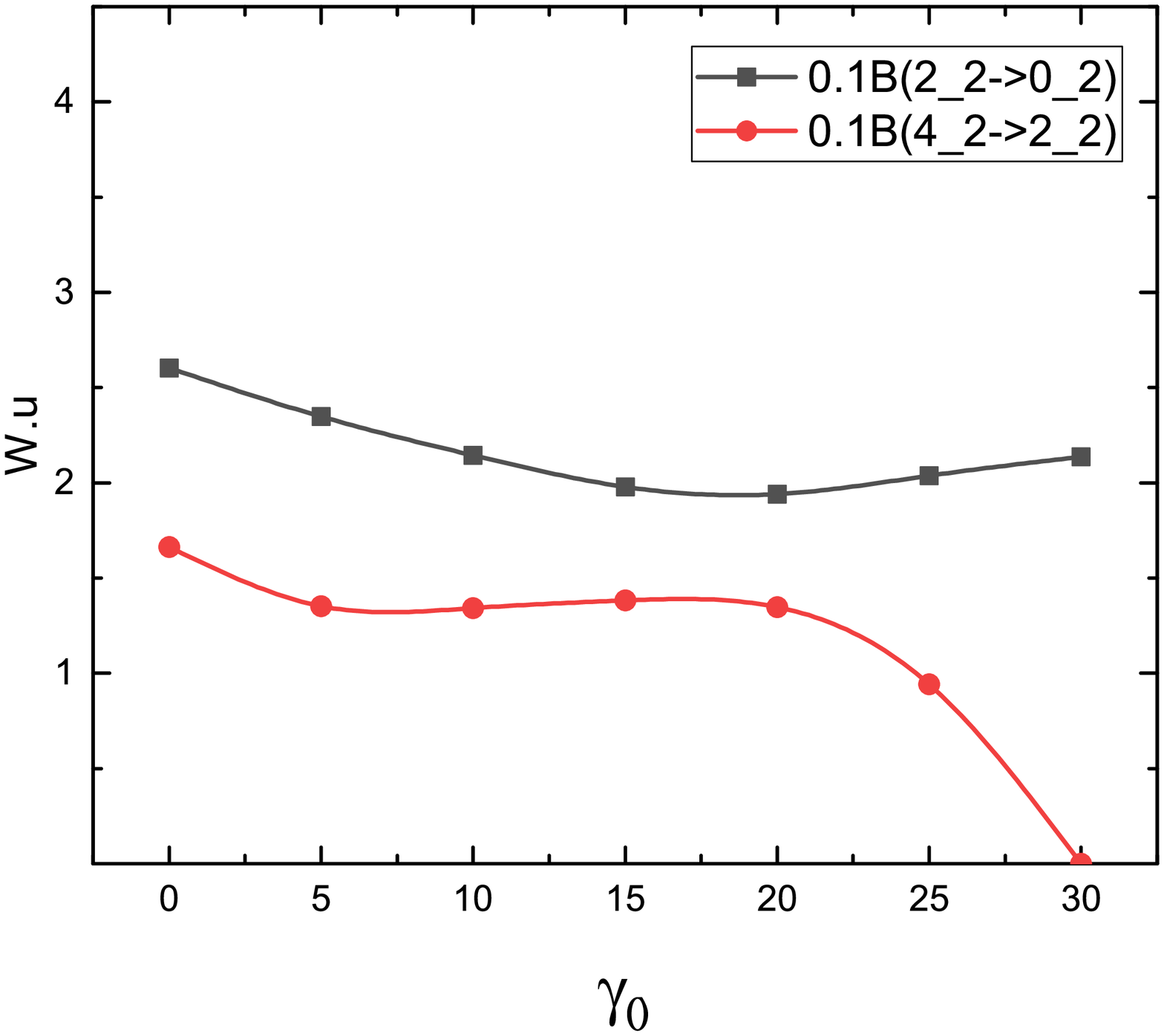}
\includegraphics[width=0.32\textwidth]{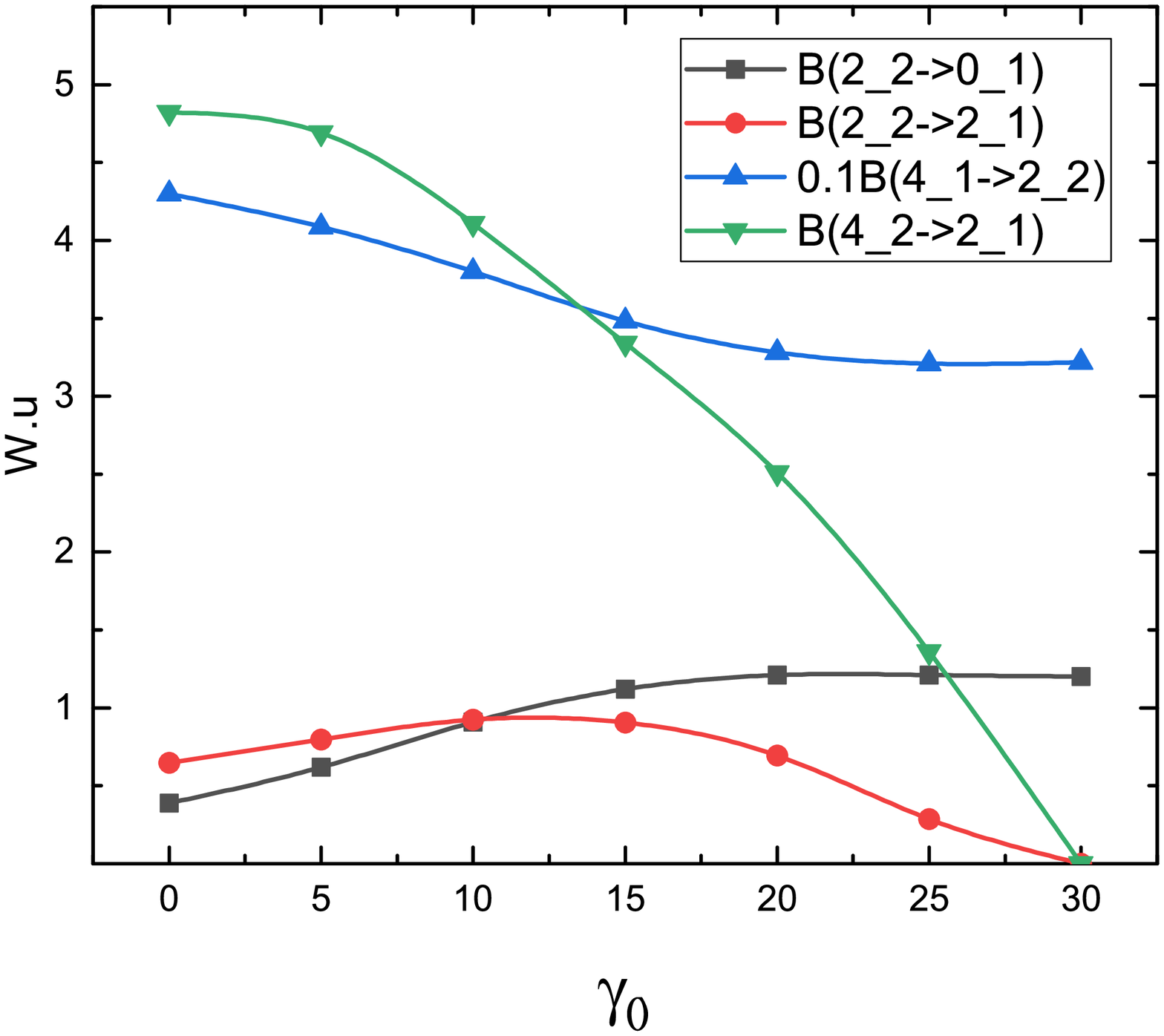}
%\end{center}
\caption{Dependence of some reduced transition probabilities on the rotation angle $\gamma_0$.}
\end{figure}
\begin{figure}[t]
\centering
\includegraphics[width=1\textwidth]{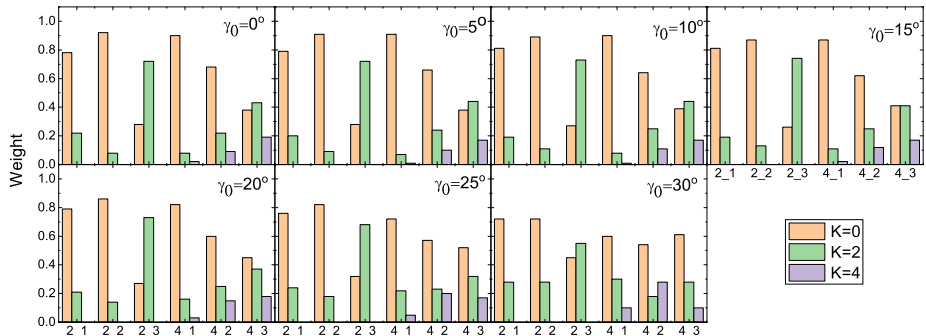}
%\end{center}
\caption{Dependence of the weights of the components on the rotation angle $\gamma_0$.}
\end{figure}
\begin{table}[h!]
\centering
\caption{Calculated energy levels depending on the rotation angle $\gamma_0$, experimental data are taken from \cite{nndc}. }
\label{tab1}
\begin{tabular}{c|c|c|c|c|c|c|c||c}
\hline
$\gamma_0$, deg	&	0&	5	&	10	&	15	&	20	&	25	&	30	&	exp	\\
\hline
$E(0^+_2)$	&	1.58	&	1.38	&	1.30	&	1.24	&	1.20	&	1.17	&	1.17&	1.58 \\
$E(0^+_3)$	&	2.44	&	2.22	&	2.15	&	2.12	&	2.11	&	2.12	&	2.14&	2.70 \\
$E(2^+_1)$	&	1.72	&	1.59	&	1.56	&	1.54	&	1.52	&	1.51	&	1.51&	1.75 \\
$E(2^+_2)$	&	2.24	&	2.00	&	1.93	&	1.90	&	1.89	&	1.90	&	1.91&	2.23 \\
$E(2^+_3)$	&	2.97	&	2.69	&	2.56	&	2.43	&	2.30	&	2.21	&	2.18&	2.67 \\
$E(3^+_1)$	&	4.21	&	3.76	&	3.56	&	3.39	&	3.26	&	3.18	&	3.15&	2.44 \\
$E(4^+_1)$	&	2.98	&	2.69	&	2.60	&	2.55	&	2.53	&	2.53	&	2.53&	2.86 \\
$E(4^+_2)$	&	3.45	&	3.20	&	3.16	&	3.14	&	3.13	&	3.13	&	3.14&	3.08 \\
$E(4^+_3)$	&	4.17	&	3.76	&	3.59	&	3.44	&	3.31	&	3.22	&	3.18&	3.18 \\
\hline
\end{tabular}
\end{table}

For the transitions between states lying in deformed well and between states belonging to different regions of deformation, the situation is more complicated. In Fig. 2 we see an exceptional behaviour of the transitions from the $4^+_2$ excited state, namely,  $B(E2; 4^+_2\rightarrow2 ^ +_1) $ and $B(E2; 4^+_2\rightarrow2 ^ +_2) $.

We see that the strength of these transitions decreases almost to zero with $\gamma_0$ approaching 30$^\circ$. Analyzing the structure of the wave function of $4^+_2$ state we see that for $\gamma_0\geqslant15^\circ$ component of the wave function with $K=4$ starts rapidly to grow. Thus the contribution from the first term of the quadrupole operator (\ref{Q2operator}) connecting the components with $\Delta K = 0$ decreases, while the contribution (which has a different sign) of the second term $Q^{coll}_{2 \mu}$ ($\Delta K = 2$), increases.

An interesting behaviour is demonstrated by the transition probability between $2 ^ +_2$ states: $B(E2; 2^+_2\rightarrow2 ^ +_1) $. We see that this matrix element decreases to zero with $\gamma_0$ increase. Again, this can be explained by the fact that, due to the increase of $K=2$ components of corresponding wave functions, $\Delta K = 2$ part of $Q^{coll}_{2 \mu}$ operator tends to cancel $\Delta K = 0$ part. The experimental value of $B(E2; 2^+_2\rightarrow2 ^ +_1)=2.8$ W.u. Therefore, to describe this probability correctly, only slight degree of triaxiality of potential energy can be assumed.

\section{Conclusion}
The influence of nonaxiality on the description of experimental data on $^{96}$Zr is investigated. In particular, the dependence of the energy levels and reduced probabilities of quadrupole transitions on the rotation angle of the potential surface is analyzed.  The main role on the behavior of the probabilities of quadrupole transitions is played by the relative weights of the components with different value of projection $K$  of angular momentum on the symmetry axis. As the angle of rotation increases, the influence of the component with $K\neq0$ increases.
This effect leads to the decrease of the probabilities of the transition between the states associated with different deformations. Therefore, one can conclude that in order to describe the experimental data on shape coexistence in $^{96}$Zr the deformed minimum of the  potential energy should be  axially-symmetric or have only slight degree of triaxiliaty. This results supports the assumptions used  in our previous papers \cite{Mardyban,Sazonov}.

\end{document}